\documentstyle[floats,multicol,aps,epsf,prl]{revtex}

\begin{document}
\draft
\twocolumn[\hsize\textwidth\columnwidth\hsize\csname @twocolumnfalse\endcsname
\def\btt#1{{\tt$\backslash$#1}}
\title{ AC Dynamics and Metastability  of a Flux-Line Lattice}
\author{W. Henderson  and E.Y. Andrei}
\address{Department of Physics and Astronomy, Rutgers University 
Piscataway, NJ 08855}
\author{M.J. Higgins and S. Bhattacharya}
\address{NEC Research Institute , 4 Independence Way, Princeton, 
	 New Jersey 08540}
\maketitle
\begin{abstract}
{We have measured the complex surface impedance of 2H-NbSe$_2$ in the
mixed state over a wide range of magnetic field(0-2T) and
frequency(10-3000MHz). A  crossover  between pinned and viscous dynamics of
flux-lines is observed at a  pinning  frequency, $\omega_p$.  
The measured $\omega _p$ is  compared to that predicted 
by  a single-particle 
 ``washboard potential'' model for the pinning
interaction. When the flux lattice is prepared in an ordered state by 
zero-field cooling or by driving it
with a large current, $\omega_p$ is in good agreement with the
predicted value.
However, when it is prepared in a metastable disordered state by field cooling,
$\omega_p$ is nearly two orders of magnitude higher than
expected indicating a complete break-down of the single-particle
model.}
 \end{abstract}
]
Pinning of the  magnetic  flux line lattice (FLL) due to material
disorder plays an important  role in 
the transport properties of   type
II superconductors\cite{blatter}. 
It can lead  to non-dissipative DC transport and to enhanced screening  of
electromagnetic  fields. 
Most studies of  pinning  have focused on  the  DC
critical current, $j_c$,   at which the FLL breaks loose  
from the pinning centers and starts moving. However, the observation of 
phenomena such as current induced
annealing\cite{kes,yaron,hnd,thesis} and
finite response times to large current pulses\cite{hnd}, 
have shown that the threshold measurements do not  completely
characterize the pinned state.  A complementary approach is to 
probe the  frequency dependence of the response 
to subcritical AC  currents which induce small oscillations about the
pinned state. This is described in a  mean-field model \cite{gr,cc} 
by the equation of motion: 
\begin{equation}
  \label{eq:gr}
  \eta\dot{u}+\kappa_pu=\Phi_0j(t)
\end{equation}
where $j(t)=je^{-i\omega t}$ is the applied current density 
(in general, thermal driving forces can be included), $u$ is
the  displacement from  equilibrium, 
 $\kappa_p$ is a restoring force constant due to pinning, 
 $\eta$  
is the  viscosity  and   $\Phi_0 =\frac {hc}
{2e}$ the flux quantum.  At a characteristic  pinning frequency, 
$\omega _p = \kappa _p /\eta $ there is a crossover from a low
frequency regime where the FLL is  pinned and the response
non-dissipative to a high frequency regime of free flux motion 
with viscous  response.  Such a  crossover was indeed  observed in 
 early AC resistivity measurements of alloy superconductors\cite{gr},
 but  recent results in  high $T_c$ samples were not 
 compatible with mean-field behavior \cite{wu}. 

\par In this Letter we report on   swept frequency measurements\cite{and}
(10-3000MHz) of 
the surface impedance in the mixed state of 
the low $T_c$  superconductor 2H-NbSe$_2$. The results are  in  good
agreement  with mean-field calculations of the electromagnetic
response\cite{cc}
  which  include contributions from Cooper pairs, the normal fluid,
flux flow and pinning. A distinct crossover frequency
is observed separating the regimes of pinned and viscous
response.  We find that $\omega_p$  
exhibits a striking sensitivity   to  the state of the FLL: for the
same field and temperature the FLL can have pinning frequencies which
differ by as much as two 
orders of magnitude, depending on how the FLL is prepared. 
This surprising  result is interpreted   in  light of pronounced
metastability effects that were observed in 
DC and pulsed current measurements \cite{hnd,thesis}.  If the FLL is
prepared  by zero field cooling to $T<T_c$ and then
applying a field, H, the FLL enters one of two  stable states. 
 For (H,T) below a well-defined transition line,
$T_m(H)$, which is marked by a  jump in  $j_c$, the state is ordered 
and weakly pinned, 
while above $T_m(H)$  it is more strongly pinned and disordered. However, 
cooling  the FLL  from above $T_c$ in a constant 
field   always results in  a  disordered FLL. 
 Below $T_m(H)$, the disordered FLL is metastable: it  can be annealed 
 into the  ordered state by driving it with a current that exceeds
$j_c$, by  changing the field  or by mechanical shock 
(after annealing  $j_c$ is up to six times lower).  The ordered 
state is stable  against such variations. 
 Our present results show the dramatic differences in the 
AC dynamics of the two states below $T_m(H)$: in the ordered  state the
FLL has a relatively low $\omega _p$  which is consistent with
the measured $j_c$ in a single-particle  washboard model. By contrast, 
 in the metastable disordered state,  
$\omega _p$  is nearly two orders of magnitude
higher  than predicted by this model, indicating the
existence of  complex underlying  dynamics. 

The experimentally accessible quantity 
 at high frequencies (where 
the AC   fields are shielded out of bulk samples) is the surface
impedance, $Z_s=R_s-iX_s$ \cite{kitt}. $R_s$ and $X_s$ are the surface
resistance and reactance. 
In the mixed state, $Z_s$ can be expressed in terms of a complex
penetration depth, $\Lambda $, by
$Z_s=-i\mu_0\omega\Lambda$. $\Lambda$ is determined  by
the response of Cooper pairs, the
normal fluid and flux-lines\cite{cc}:
\begin{equation}
\label{eq:r_a}
\Lambda  =\left(\frac {\lambda ^2+i\delta ^2_v/2}
 {1 - 2i\lambda ^2/\delta ^2_{nf}}\right )^{1/2}.
\end{equation}
Here   $\lambda$  is the London penetration depth, 
$\delta _{nf}$ is  the normal
fluid skin depth, \mbox{$\delta _v=(2\rho _v/\mu_0\omega )^{1/2}$} is the
 flux-line skin depth and  $\rho _v$ is the  flux-line
resistivity. Our results are expressed in  terms of  the reduced  
variables,  \mbox{$r_s=R_s$(T)/$R_s$(T$_c)=Im(\frac {2\Lambda}
{\delta _n}$)} and \mbox{$x_s=X_s$(T)/$X_s$(T$_c)=Re(\frac{2\Lambda
}{\delta _n}$)}, which eliminate a trivial $\omega^{1/2}$ 
dependence, due to the skin effect. 
The normal state skin depth, $\delta_n$, is almost  independent of H
and T in our samples. This was determined from  measurements of  the
normal state resistivity, $\rho_n$.  
Most results described here are in a range of H,T, and $\omega$  
where  
$\lambda\ll\delta_v,\delta_{nf}$ and the  response 
is dominated by the flux-lines:
\begin{equation}
\label{eq:z_v}
r_s-ix_s=(-2i\rho_v/\rho_n)^{1/2}.
\end{equation}
 If we assume 
the Bardeen-Stephen value for the viscosity, $\eta=\Phi_0
H_{c2}$/$\rho_n$,  with $H_{c2}$ the upper critical field, 
then in the model
described by (\ref{eq:gr}), the flux-line resistivity is: 
\begin{equation}
\label{eq:rv}
\rho_v(\omega) = \rho_{v1}(\omega)-i\rho_{v2}(\omega) =
\frac{\omega^2-i\omega\omega_p}{\omega^2+\omega_p^2}\frac
{H}{H_{c2}}\rho_n.
\end{equation}
In the free  flux flow limit ($\omega \gg \omega _p$), the
response  is  further simplified and reduces to that of 
 a  normal metal with a purely real resistivity,
$\rho_nH/H_{c2}(T)$:

\begin{equation}
\label{eq:r_x}
\! r_s(T,H)\!=\!x_s(T,H)\!=\!
\!\begin{array}{ll}(H/H_{c2}(T))^{\frac{1}{2}} &
\quad \mbox{$T\leq T_c(H)$} 
\end{array}.
\end{equation} 

Measurements were done on two single crystal samples.
For sample 1 ($\sim$2.5 x 2.5 x .1mm)
 $T_c=$ 7.2K, $\Delta T_c=$80mK,  and the residual resistivity ratio,
RRR=23.  For sample 2 ($\sim$4 x 4 x .025mm) $T_c$ = 5.85K, $\Delta 
T_c$=80mK and  
RRR =9. The parameters indicate that sample 2 is relatively impure,
and that both  have good homogeneity. The field was 
along the sample's c-axis, and the AC current was in the a-b plane. We will
focus on data for sample 2. 

\begin{figure}[btp]
\epsfxsize=3.5in
\epsfbox{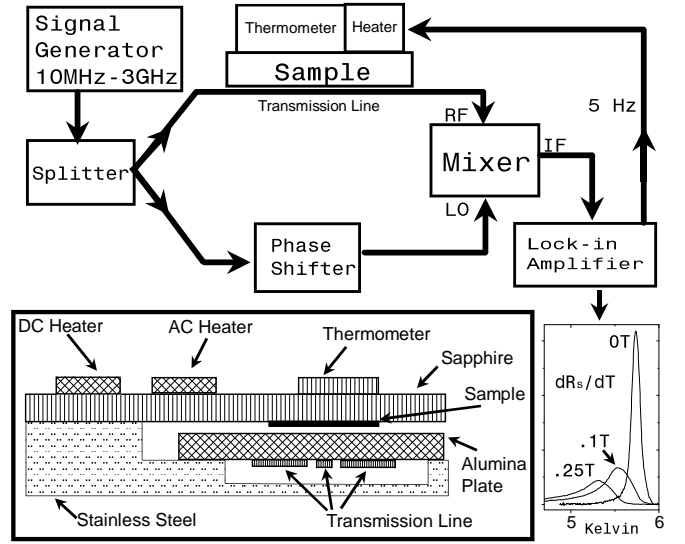} 
\protect\caption{Upper Panel: schematic diagram of the measurement
technique. Lower panel: cross sectional view of the sample holder and
transmission line.}
\label{fig:fig_rf1}
\end{figure}

The experimental setup \cite{thesis}  is shown in Fig.
\ref{fig:fig_rf1}. The sample is 
mounted on a sapphire plate, which is attached to a stainless steel
holder inside a  vacuum can. A 50$\Omega$ coplanar 
transmission line consisting of three Cu strips 
(one central conductor between two ground planes) 
on an alumina substrate is placed  close to the
sample. The sample's surface resistance  (reactance) introduces 
a slight change ($\leq 10^{-4}$)  in the magnitude (phase)  of
the signal propagating along the transmission  line.
 This effect is isolated from a large background by modulating the 
sample temperature  at  a low frequency ($\sim$5-10Hz) with a
resistive heater.   The  mixer puts out a signal at the modulation
frequency, which is proportional to
$cos(\phi)dR_s/dT+sin(\phi)dX_s/dT$, and 
is detected with  a lock-in amplifier.   By tuning $\phi$,  the phase
difference between the LO and RF ports, the derivative with respect to
temperature of $R_s$ and $X_s$ can  
be measured separately. All the  data were taken in the linear
regime: the RF currents in the sample were 
$\sim$1mA ($j\sim 1A/cm^2 \ll j_c$) and the 
 amplitude of the flux-line motion ($<$.1\AA \ at 10MHz) was
negligible compared to their spacing.

\begin{figure}[btp]
\epsfxsize=3.5in
\epsfbox{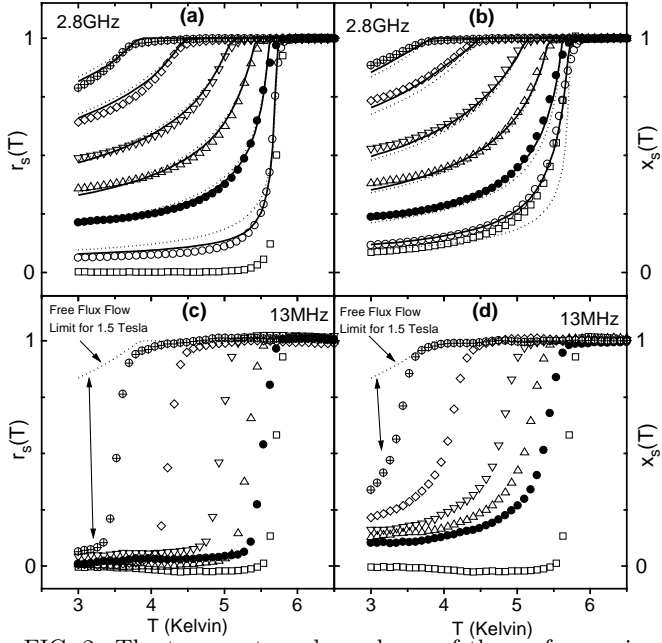} 
\protect\caption{The temperature dependence of the surface resistance
and surface reactance at 2.8GHz (a and b) and 13MHz (c and d) for 0
field ($\Box$), .02T ($\bigcirc$), .1T ($\bullet$), .25T
($\bigtriangleup$), .5T 
($\bigtriangledown$), 1T ($\Diamond$), and 1.5T ($\otimes)$. The FLL
was field cooled in all cases. The dotted
lines were calculated using (\ref{eq:r_x}); the solid lines were
calculated using (\ref{eq:r_a}). The double arrows show the deviation
from free flux flow behavior.}
\label{fig:fig_rf2}
\end{figure}

We first present the results in the  disordered  state. 
The FLL was prepared in this state by cooling 
 from above $T_c$ to 3K in  a constant field.
In Fig. \ref{fig:fig_rf2} we show the T dependence of $r_s$
and $x_s$. The data was obtained by slowly ramping T 
 as  $dR_s/dT$ or  $dX_s/dT$ was measured. 
$r_s$ was obtained from the raw data in the following manner.
The zero field data were integrated, with  the  integration constant 
fixed by assuming that $R_s$=0 for $T\ll T_c$. 
The  integrated data were scaled
 to be equal to 1 at $T_c$. The value of  the  scaling factor 
 depends on 
the  coupling between sample and transmission line which is 
 independent of H and therefore  the same scaling factor was used 
for the finite field data.  The finite field integration constants
were set by using the fact that $Z_s$ is independent of field in the
normal state. 
The same basic procedure was used for  the reactance 
data, but  determining the integration constants was complicated by the
fact that even at low temperatures and zero field,
 $X_s$ has a finite value $\mu_0\omega\lambda$  and the absolute
value of  $\lambda$  was
not  measured directly. The necessary  parameter 
 $x_0\equiv x_s(3K,0T)=2\lambda(3K,0T)/\delta_n$,  was determined
by the method described below. 

 In Fig. \ref{fig:fig_rf2}a the measured    $r_s(T)$ at 2.8
GHz is  compared to the calculated   values.   
For fields $\geq$.1T, the approximate expression
(\ref{eq:r_x}) is in good  agreement with the data. This calculation
had no adjustable parameters since  $H_{c2}(T)$ was obtained from DC
measurements of  $T_c(H)$.  The agreement indicates 
that for this frequency  the FLL is in the free flux flow regime 
and consequently the pinning frequency for the disordered state
$\omega ^d_p\ll$2.8GHz.  At lower fields,
$\delta_v\sim \lambda$ and the approximate expression is no longer
adequate. Fitting the data at .01, .02, and .05T to the full
expression (\ref{eq:r_a}), with
$x_0$ as an adjustable parameter, we obtain 
 $x_0\sim$.09 at 2.8GHz. This value was used to process the reactance 
data  shown in \mbox{Fig. \ref{fig:fig_rf2}b}. We see that the full
expression is in good agreement with the data at all fields for both
$r_s$ and $x_s$. (A small contribution from pinning, determined by the
measured pinning frequencies, was included in these calculations. At the
higher fields, this is the main source of the slight difference between
the calculations using (\ref{eq:r_a}) and (\ref{eq:r_x})). 
 Since $x_0=\frac{2\lambda (3K, 0T)}{\delta _n}$ and  $\delta_n$ 
depends only on $\rho_n$, which was measured,  the value of $x_0$ 
can be used to  determine $\lambda$. From the data  at several
  frequencies in the GHz range,
we estimate  $\lambda\sim$1200-1400\AA \ in sample 2 and
$\sim$1000-1200\AA \ in sample 1. Previous estimates are scattered
over the range 700-2500\AA \ \cite{lamb}. 

\begin{figure}[btp]
\epsfxsize=3.2in
\epsfbox{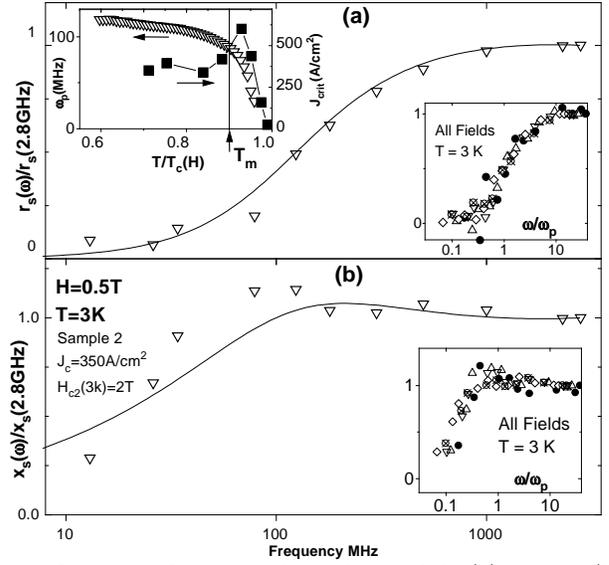} 
\protect\caption{The frequency dependence of $R_s$ (a) and $X_s$ (b) at
3K and 0.5T. (a), upper inset: field and temperature dependence
of the pinning frequency. Lower insets: the collapse of the data for
all fields. Symbols are the same as in  Fig. \ref{fig:fig_rf2}.
\label{fig:fig_rf3}}
\end{figure}

In Figs. \ref{fig:fig_rf2}c and \ref{fig:fig_rf2}d we show the results
at  13MHz. In this case,  
both $r_s$ and $x_s$ are well  below the free flux flow 
values,  indicating  that  $\omega^d _p >13$MHz. 
The frequency dependence of the response over the entire 
range,  shown  in
Fig. \ref{fig:fig_rf3},  exhibits a 
crossover from pinned to free behavior.  
The best  fit  of the data to (\ref{eq:z_v}), with $\rho_v$
given by (\ref{eq:rv}), is obtained for 
$\omega^d_p = 125$MHz.
The upper inset of  Fig. \ref{fig:fig_rf3}a, shows the temperature dependence of
$\omega^d_p$ and $j_c$ at .5 T. 
The  lower insets in Fig. 3 show that the data for all 
fields collapse when plotted as a function of $\omega /\omega^d _p(T,H).$ This universal behavior 
demonstrates that in this state the frequency dependence of the response of the FLL 
 is completely characterized  by one parameter, $\omega^d_p$,
 as expected from (1). 

\begin{figure}[btp]
\epsfxsize=3.5in
\epsfbox{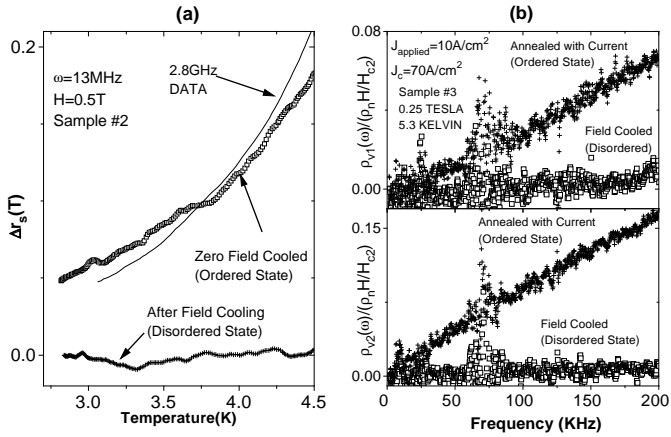} 
\protect\caption{Comparisons of the AC response for the disordered
and ordered states below $T_m$ 
(a) $\Delta r_s(T)$ at 13MHz.
(b) The frequency dependence of the complex
resistivity. The  noise near 75KHz is instrumental.} 
\label{fig:fig_rf4}
\end{figure}

We now turn to the  response in the ordered state. When the FLL is
prepared 
in this state and then   heated above
$T_m$, it undergoes an order to
disorder transition that is \emph{irreversible in temperature}
\cite{hnd}. Thus, $dR_s/dT$ and $dX_s/dT$ cannot be obtained with this
technique over the entire temperature range and it is not 
possible to set the integration constants (since this requires
data up to $T_c$). Although 
 the absolute values of $r_s$ and $x_s$ cannot be determined, 
 the relative changes, 
 $\Delta r_s(T)$ and $\Delta x_s(T)$, are
still accessible 
for $T<T_m$. At frequencies in the GHz range, the results are
independent of how the FLL is prepared.  In Fig. \ref{fig:fig_rf4}a,
we plot  $\Delta r_s(T)$  at 13MHz in both  states. 
The data in the disordered  state is
essentially constant in temperature, consistent  with the high pinning
frequency of this state. In the ordered state  
 however the data has a substantial slope which is close to 
its value at 2.8GHz, indicating that the pinning frequency  in this state
 $\omega^o _p<$13MHz. 

Since the value of $\omega^o _{p}$  appears to be in the MHz range, 
it can be measured by more  conventional means. 
At these frequencies the sample thickness  is much less than 
$\delta_{v}$, so that the skin effect is negligible
and $\rho _v$  could   be accessed 
directly with a  4-lead  technique. The sample used 
for these measurements  was from the same batch as sample 2 
and had nearly  identical parameters. A background signal was determined from
zero field  measurements at  $T\ll T_c$ and subtracted from the
data. 
In Fig. \ref{fig:fig_rf4}b we see
that in  the disordered state  both $\rho _{v1}$ and $\rho _{v2}$ 
are vanishingly small over the entire frequency range, 0-200kHz,
indicating that  
$\omega^d _p \gg $200kHz, in accord with the previous result. 
By contrast,  $\rho _{v1}$ and $\rho _{v2}$ are finite in the
ordered state and their  frequency  dependence  gives  
, $\omega ^o _p\sim$1MHz. 

To interpret the results, we use  a model where  the FLL is treated as 
 a single particle  moving in a washboard  
potential: $V(u)=V_0(1-cos(\frac{2\pi u}{r_p}))$. $u$ is the
displacement from equilibrium  and $r_p$ 
is a characteristic length scale for the interaction of the FLL with
the pinning sites. In this model,   the critical
current is  $j_c=\frac{2\pi V_0} {r_p\Phi _0}$ and the   restoring force
constant is  $\kappa_p =V_0(\frac {2\pi}{r_p})^2 =j_c \frac {2\pi\Phi_o} {r_p}$
leading to: 

\begin{equation}
\label{eq:omega_p_o}
\omega_p=\frac{\kappa_p}{\eta }=\frac {2\pi}{r_p}\frac{j_c\rho _n
}{H_{c2}}.   
\end{equation}

For the ordered state  data shown in 
\mbox{Fig. \ref{fig:fig_rf4}b}, $j_c$ = 70A/cm$^2$. Choosing 
 $r_p$=$a_0$ (where $a_0\cong(\Phi_0/H)^{1/2}$ is the flux
line spacing), as is usually done, 
gives $\omega^o_p$=1.4 MHz, in rough agreement with the data. 
In  the disordered state ($j_c(3K,0.5T)$=
350A/cm$^2$), the washboard model gives $\omega_p$= 2.6 MHz,
which is 45 times lower than the  observed
value. 
This discrepancy is similar everywhere in the H-T plane except 
above T$_m(H)$ 
(where the disordered state is stable), in which case it is
much smaller.  
The washboard model can give a  higher $\omega _p$ 
if  $r_p$ is reduced.
But even using the smallest length scale in the problem, 
the coherence length,  and taking into account the structure of the
flux-line core\cite{blatter} leads to values 
 that are not much larger: $\omega_p$(3K,.5 T)=4.7 MHz. 
In addition, the washboard model cannot account for the qualitative
difference in the  temperature dependences 
of $\omega_p$ and $j_c$. 
Thus, this model
successfully describes the connection between $j_c$ and $\omega _p$ in
the ordered state, but fails to do so in the metastable disordered  state. 
 This  is due to the  single particle treatment of the FLL  in  which  
 both $j_c$ and $\omega _p$
 are determined by the shape of a pinning potential. In this model,
$j_c$ is the 
current to drive the entire FLL out of a potential. 
 However, the metastable state reorders as it depins 
and $j_c$ for this state may actually be the current at which the FLL
becomes unstable to rearrangements  which create
continuous channels of mobile flux 
lines. Evidence for channel formation and growth was previously
observed with DC\cite{hnd,thesis,kap} and pulsed current
measurements\cite{hnd}, as well as in computer
simulations\cite{berl}. However, in the AC measurements, no
rearrangements can take place because of the low driving currents and
the short time scales.  

In summary, the AC response in both states exhibits  a crossover 
between pinned and free behavior  which can be  described  by a simple
equation of motion. In the ordered state 
 the connection between  $\omega _p$ and $j_c$ is accurately predicted by
a single particle  washboard model, but for the  metastable disordered state
  the model breaks down. 
The two states discussed here are possible
candidates for the proposed vortex glass (disordered) and pinned
lattice or Bragg glass (ordered)\cite{huse}. Our results indicate that
each of these states can be supercooled or superheated across T$_m$ into the stability region of the other\cite{hnd,thesis}. In this picture,
the metastable disordered state corresponds  to a   supercooled vortex glass. 

We  thank N. Andrei,   D.A.  Huse and R. Walstedt for useful
comments. Supported by  NSF-DMR-9401561.

\end{document}